\documentclass[conference]{IEEEtran}

\usepackage{cite, makecell, subcaption}
\usepackage{amsmath,amssymb,amsfonts}
\usepackage{algorithmic}
\usepackage{graphicx}
\usepackage{textcomp}
\usepackage{xcolor, rotating}
\usepackage{balance}
\usepackage{multirow}
\usepackage{gensymb}
\usepackage{array,booktabs}
\usepackage{longtable,tabularx,tabulary}
\usepackage{lipsum}
\usepackage{amssymb,amsmath,epsfig,latexsym, bm}
\usepackage{hyperref}
\hypersetup{colorlinks, pdfa=true, breaklinks, citecolor=blue, urlcolor=black, linkcolor=black}
\usepackage{hyphenat}

\def\BibTeX{{\rm B\kern-.05em{\sc i\kern-.025em b}\kern-.08em
T\kern-.1667em\lower.7ex\hbox{E}\kern-.125emX}}

\newcommand*{\email}[1]{\normalsize\texttt{\href{mailto:#1}{#1}}\par}

\DeclareRobustCommand\onedot{\futurelet\@let@token\@onedot}
\def\@onedot{\ifx\@let@token.\else.\null\fi\xspace}

\makeatother

\begin{document}

\title{Quantum Hybrid Support Vector Machines for Stress Detection in Older Adults}

\thanks{978-1-6654-6658-5/22/\$31.00 ©2022 IEEE}

% uncomment before sending to blind review
%\author{\vspace{0.85in}}

% comment before sending to blind review
% TODO: Add last author
\author{
    \IEEEauthorblockN{%
         Md. Saif Hassan Onim$^1$,
         Travis S. Humble$^2$ and
         Himanshu Thapliyal$^1$
                    }

    \IEEEauthorblockA{%
        $^{1}$Department of Electrical Engineering and Computer Science,\\
        University of Tennessee, Knoxville, TN, USA\\
        $^2$Quantum Science Center, Oak Ridge National Laboratory, Oak Ridge, TN, USA\\}
    
\email{monim@vols.utk.edu},
\email{humblets@ornl.gov},
\email{hthapliyal@utk.edu}            
    }
\maketitle

\begin{abstract}
Stress can increase the possibility of cognitive impairment and decrease the quality of life in older adults. Smart healthcare can deploy quantum machine learning to enable preventive and diagnostic support. This work introduces a unique technique to address stress detection as an anomaly detection problem that uses quantum hybrid support vector machines. With the help of a wearable smartwatch, we mapped baseline sensor reading as normal data and stressed sensor reading as anomaly data using cortisol concentration as the ground truth. We have used quantum computing techniques to explore the complex feature spaces with kernel-based preprocessing. We illustrate the usefulness of our method by doing experimental validation on 40 older adults with the help of the TSST protocol. Our findings highlight that using a limited number of features, quantum machine learning provides improved accuracy compared to classical methods. We also observed that the recall value using quantum machine learning is higher compared to the classical method. The higher recall value illustrates the potential of quantum machine learning in healthcare, as missing anomalies could result in delayed diagnostics or treatment.

\end{abstract}


\section{Introduction}
\label{intro}

Excessive stress can harm a person's mental health leading to various medical conditions. It can raise the risk of mental and physical disorders, such as immune system issues, cardiovascular and metabolic system impairment and more. Besides, stress is connected to cognitive decline, which over time could increase the onset of Alzheimer's Disease in older adults, as well as physiological irregularities. Multiple studies suggest that experiencing a longer period of stressful situations can contribute to the deterioration of mental and physical health, as well as mortality~\cite{Mcewen2008, Cohen2007}. Thus, early detection of stress is necessary for successful intervention. Recent advances in machine learning and quantum computing (QC) have created new opportunities to solve such problems with artificial intelligence (AI)~\cite{qml, ciliberto2018}. Quantum machine learning (QML) is a promising field generating active research across multiple domains. Researchers have increasingly used various quantum algorithms for real-life healthcare applications. 
In this domain, quantum machine learning has grown as an emerging AI framework. It attracts interest from both academic and medical communities. Researchers made significant experimental advances in classifying and diagnosing stress. There have been recent advancements made in the development of quantum support vector machines (QSVM), Quantum Neural Networks, variational quantum classifiers (VQC), error reduction tactics, and advanced pre-processing methods~\cite{ Schuld2019, moll2018, rebentrost2014, chen2020, cultice2024}. QML uses the encoded classical dataset which is transformed into quantum states. There are several benefits of encoding data into quantum states. It helps perform quantum computation with classical data. For linear classifiers, kernel-based methods have also been effective in providing dataset uniformity. Similar to classical machine learning, researchers are currently working to create reliable quantum algorithms that are suited to classifying problems unique to mental health such as stress detection~\cite{padha2023, ullah2024}. Although, there have been multiple experiments on classical algorithms that can detect and predict stress~\cite{casdoa, onim2024utilizing, rhodus2024utilization, onim2024predict}. In that regard, hybrid QC and ML algorithms have been applied to multiple datasets~\cite{hai2020, hai2021, PhysioBank, ephnogram}. These algorithms combine classical algorithms with the benefits of quantum computing approaches. Again, the application of such framework for older adults is under-explored~\cite{casdoa,onim2024utilizing, nath2022} while there are a lot of works focusing the young adults~\cite{Parent2019, taylor2017}. Thus, developing innovative machine learning approaches that take advantage of QC's specific features for healthcare applications is a timely necessity. Their performances also need to be evaluated for real-life scenarios that can justify the viability of such a framework for older adults.

To address the issue, we experiment with the kernel-based Quantum Hybrid Support Vector Machine (QHSVM) on a real-life stress dataset of older adults in a constrained environment. Our proposed hybrid framework looks at the stress detection problem as an anomaly detection problem. The inner product kernel finds the fidelity between features to capture the inherent pattern that distinguishes the anomaly from the baseline. Finally, we show a performance analysis with experimental results on quantum simulation.

\section{Proposed Method}
\label{Methodology}
\begin{figure}[htbp]
\centering
\includegraphics[width=.6\columnwidth]{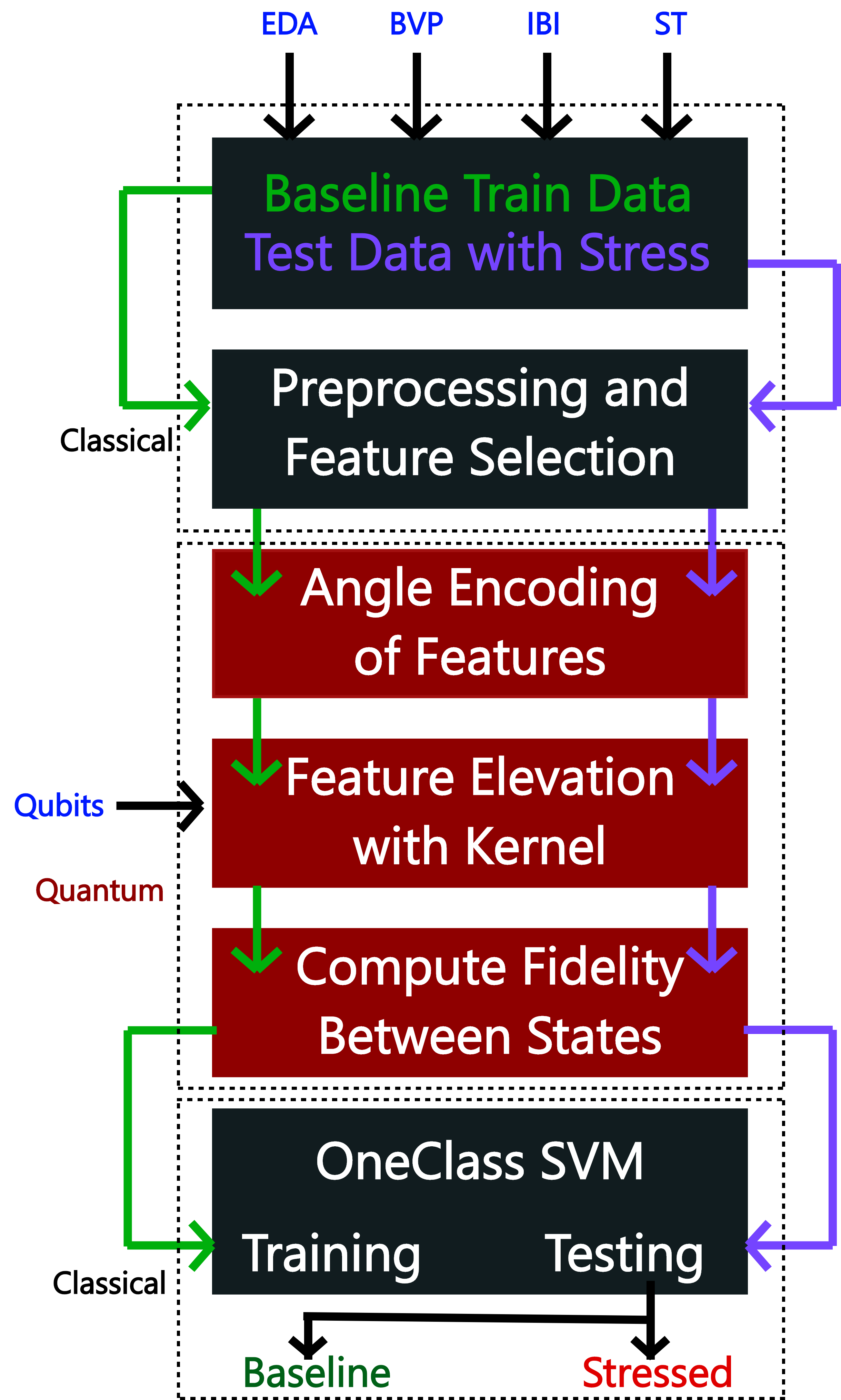}
\caption{Pipeline of the hybrid-quantum SVM anomaly detection}
\label{fig:workflow}
\end{figure}

The procedure for this hybrid SVM method consists of the following stages: first, we perform feature selection and pre-processing in the classical machine. Then, we compute the fidelity between features with the inner product of the probability taken from the quantum kernel circuit. Finally, use this information to train and classify a OneClass SVM in the classical machine. The whole process is shown in Figure~\ref{fig:workflow}.

% Figure here: Overall Flowchart

\subsection{Dataset Collection and Feature Extraction}
\label{Dataset}

\begin{figure}[htbp]
    \centering
    \includegraphics[width=\columnwidth]{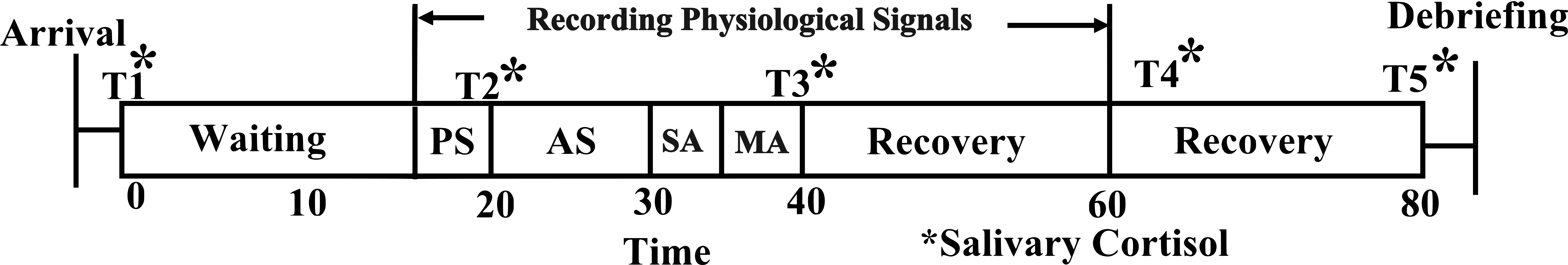}
    \caption{TSST Protocol}
    \label{fig:tsst}
\end{figure}

In this dataset, there are 60 features extracted from 4 signals (Electrodermal Activity) EDA, (Blood Volume Pulse) BVP, (Inter Beat Interval) IBI, and (Skin Temperature) ST collected from Empatica E4 wristband. There were 40 participants and the training data was captured from baseline labeled with cortisol concentration. Participants' age ranges from 60 to 80 years. This group comprised 28 female and 12 male individuals. Salivary Cortisol was collected during the (Trier Social Stress Test) TSST protocol which is a stress-inducing technique. The TSST is a well-known experimental framework that could generate stress in a naturalistic setting \cite{tsst}. Figure~\ref{fig:tsst} illustrates the TSST's procedural sequence, which includes a few distinct phases. It is divided into four stages: the waiting interval, the pre-stress phase, the stress induction period, and the recovery phase. Participants in the preliminary stage fill out demographic questionnaires. Following the waiting period, the pre-stress (PS) phase begins, during which baseline data are collected. The total duration of the waiting period and pre-stress phase is 20 minutes (T1–T2). Then the pre-stress and stress phases last 20 minutes (T2-T3). The stress-inducing session is divided into three distinct segments: an initial 10-minute period known as the anticipatory stress phase (AS), followed by a 5-minute interval that includes speech and cognitive arithmetic tasks (M). During the anticipatory stress phase (AS), participants are asked to maintain a continuous discourse on a specific topic for 5 minutes while being monitored. Then they participate in a cognitive arithmetic exercise that involves completing simple addition and subtraction calculations. However, the complexity of the cognitive activities increases with each correct answer. The trial concludes with two consecutive 20-minute recovery phases (T3–T5). Salivary samples were obtained five times during the trial, at time points T1, T2, T3, T4, and T5.

The test data contains both normal (baseline) and anomalous(stressed) data of an equal number of points making it a balanced dataset. The preprocessing stage is intended to prepare the data for use with SVM and feature selection techniques. Initially, categorical samples are replaced with numerical histogram values, which are more appropriate for machine learning techniques.

% Encoding these 60 features to a quantum machine to perform the kernel computation will be expensive both in terms of run-time and resources. With an additional qubit, the computational complexity increases quadratically. 

Encoding 60 features to quantum states to perform the kernel computation is a very computationally expensive task and the runtime will also be impractical. For each set of additional features, the complexity increases quadratically. One way to address that would be parallel computation but the increase in features also increases the circuit width keeping the circuit depth the same. This prevents the parallel computation of the kernel. Again, some less correlated features tend to throw the machine learning models off from convergence resulting in poor accuracy. Thus to accommodate the highly correlated features with reasonable complexity, we select the best 8 and 12 features with the feature selection technique. To select the best features, a decision tree is trained using all training data. The tree can separate each sample of the features based on the splitting criteria or ground truth. Later the features are scored in accordance with their ability to reduce the error or impurity. For decision Trees or Random Forest based models, the error is calculated with \textit{Gini Index} and \textit{Entropy}. For our work, we ranked the features according to their relative decrease in Gini impurity, as shown in Equation~\eqref{gini}, \eqref{dec_gini}. In this equation, $I$ represents the impurity estimated for the probability $P$ of a sample occurrence in node $N$, while $I_0$ represents the impurity calculated from the preceding node $N_0$.

\begin{equation}
    I(n) = 1-\sum P^2
    \label{gini}
\end{equation}

\begin{equation}
    \Delta I = I_0 - \left( \frac{N_{left}}{N_0}\times I_{left} + \frac{N_{right}}{N_0}\times I_{right} \right)
    \label{dec_gini}
\end{equation}

Again, before feature extraction, we randomly sample 2000 baseline data points for training and 1500 data points with a combination of baseline and stress labels for testing.

\subsection{Fidelity Kernel}

\begin{figure}[htbp]
\centering
\includegraphics[width=\columnwidth]{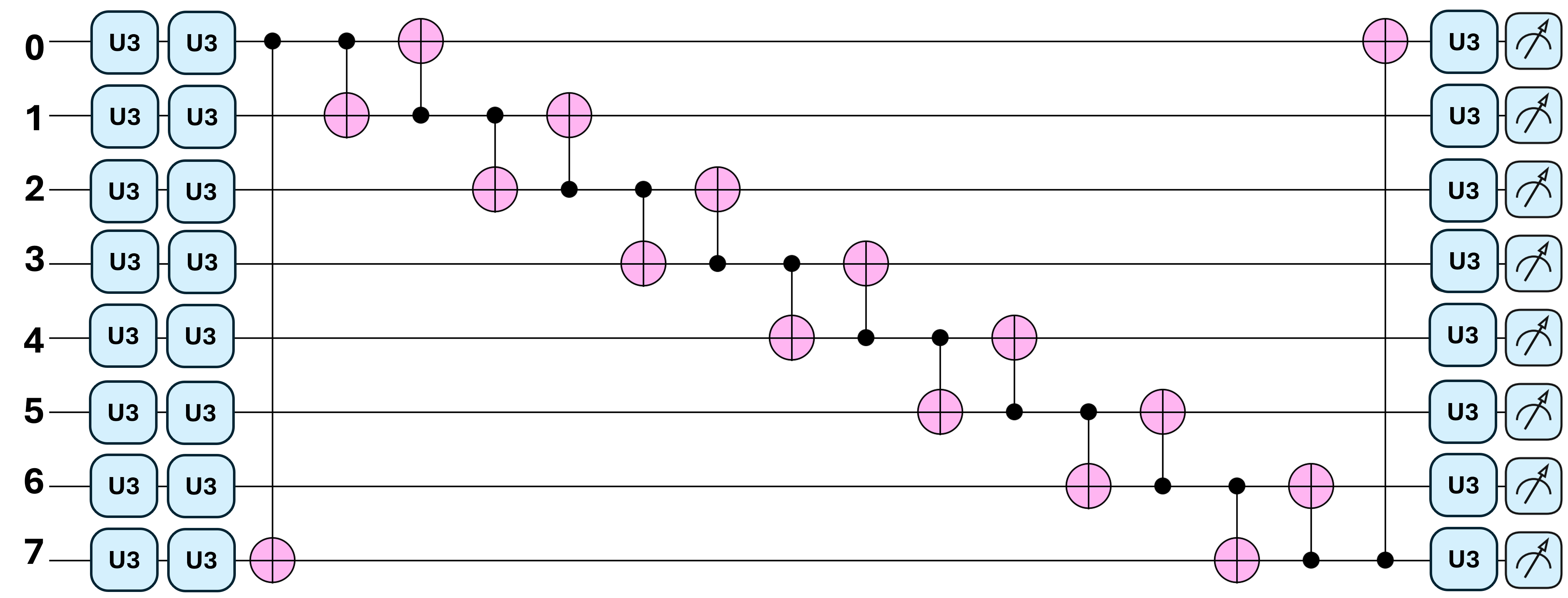}
\caption{Quantum Circuit for Feature Mapping with 8 qubits}
\label{fig:featureMap}
\end{figure}

The quantum kernel circuit is made with three steps. The first part is the encoding where each lines are initialized as a $0$ state of qubits. Here, the classical data is encoded in a way that each qubit can represent two features. This encoding scheme named Dense angle encoding was proposed by~\cite{LaRose2020}, where $N$ features can be encoded into $N/2$ qubits. Fig.~\ref{fig:featureMap} shows the kernel circuit. $U3$ is a rotation unitary consisting of a single qubit rotation that works as angle encoding. The $R_x$ gate applies a rotation about the x-axis of the Bloch sphere, mathematically defined as $R_x(\theta) = e^{-i\theta X/2}$, where $X$ is the Pauli-X matrix and $\theta$ is the angle of rotation, which can be set as proportional to $x_1$. Similarly, the $R_z$ gate rotates the qubit around the z-axis and is defined as $R_z(\phi) = e^{-i\phi Z/2}$, with $Z$ being the Pauli-Z matrix and $\phi$ proportional to $x_2$. By applying $R_x(x_1)$ followed by $R_z(x_2)$ to each qubit, the two features $x_1$ and $x_2$ are mapped to the qubit's quantum state. Afterward, the series of $CNOT$ gates entangle a feature with the rest of them. After the measurement, the probability of keeping initial states is fed into an inner-product kernel based on quantum fidelity. It is expected to measure a low fidelity and higher for the baseline. The fidelity of the same feature was normalized to $1$. The Fidelity kernel captures the high-dimensional similarities among the features. This circuit is constructed following the work of~\cite{woniak2023}.

\subsection{Training and Testing with the One-Class Support Vector Classifier}
In the training phase, the OneClass SVM finds the ideal hyperplane between classes by identifying the support vectors. In kernel-based OneClass SVM, similar data points are bounded by a decision hyperplane where the anomalies are kept out. If $\{x_r\}_{r=1}^N$ represents the data from a single class, the decision function for a One-Class SVM can be represented as Equation~\eqref{obj_func}.

\begin{equation}
    f(x) = sign\left( \sum_{r=1, s=1}^N \alpha_r \times k(x_r, x_s) + b - \rho \right)
    \label{obj_func}
\end{equation}

Here, $k(x_r, x_s)$ is the kernel function, and $x_r$ and $x_s$ are the input vectors. $\alpha_r$ are the tunable constants. $b$ is the bias term and $\rho$ is the threshold. The inner product kernel shape for training samples is $N_{train} \times N_{train}$ with a symmetry by the diagonal and the shape for the test kernel is $N_{test} \times N_{train}$. Projecting the testing kernel in a 2D space can help visualize the difference between the classical and quantum kernels.

\begin{figure}
    \begin{subfigure}[h]{0.485\linewidth}
    \fbox{\includegraphics[width=\linewidth]{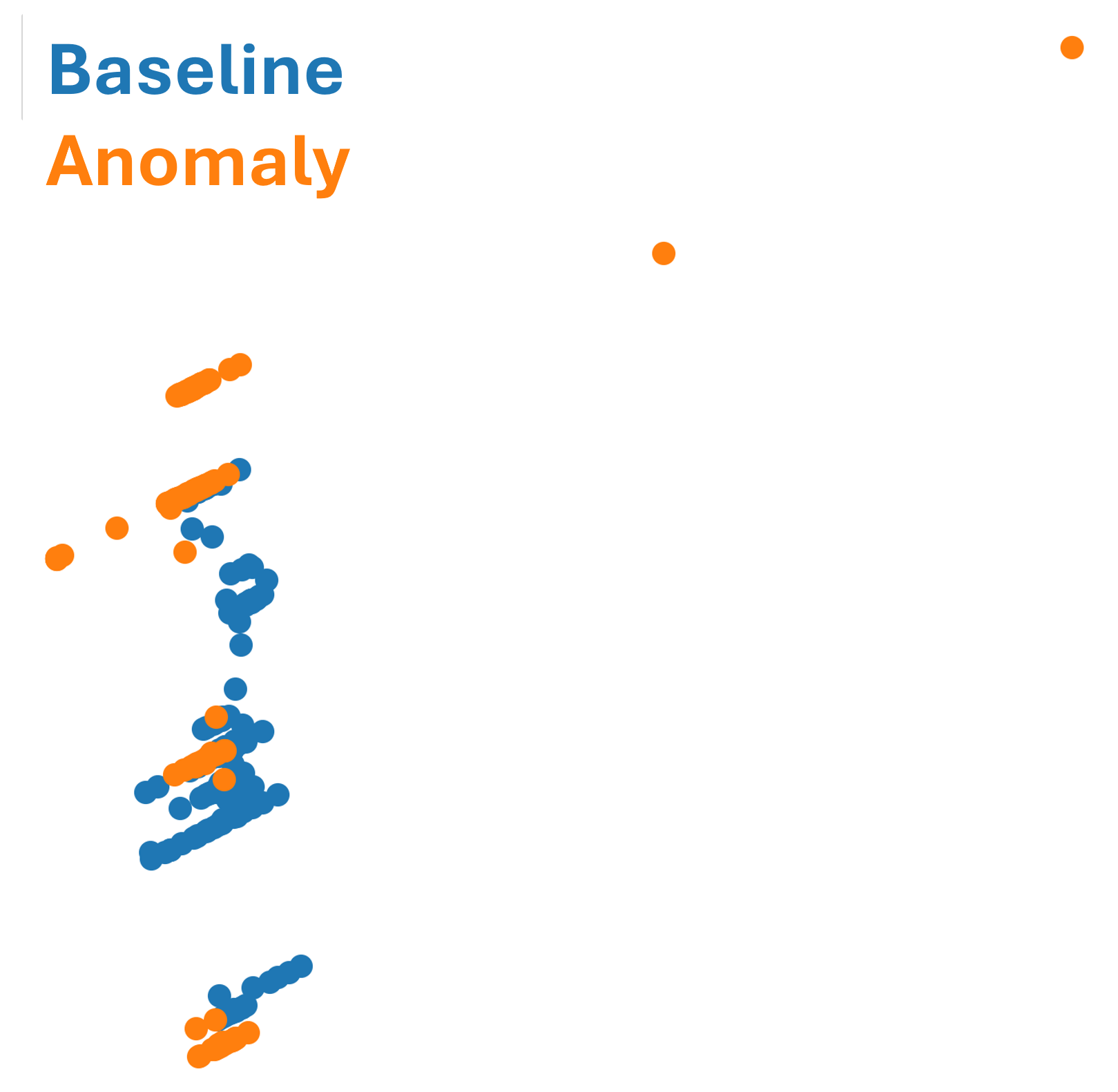}}

    \caption{Classical Kernel}
    \label{fig:ck}
    \end{subfigure}
\hfill
    \begin{subfigure}[h]{0.47\linewidth}
    \fbox{\includegraphics[width=\linewidth]{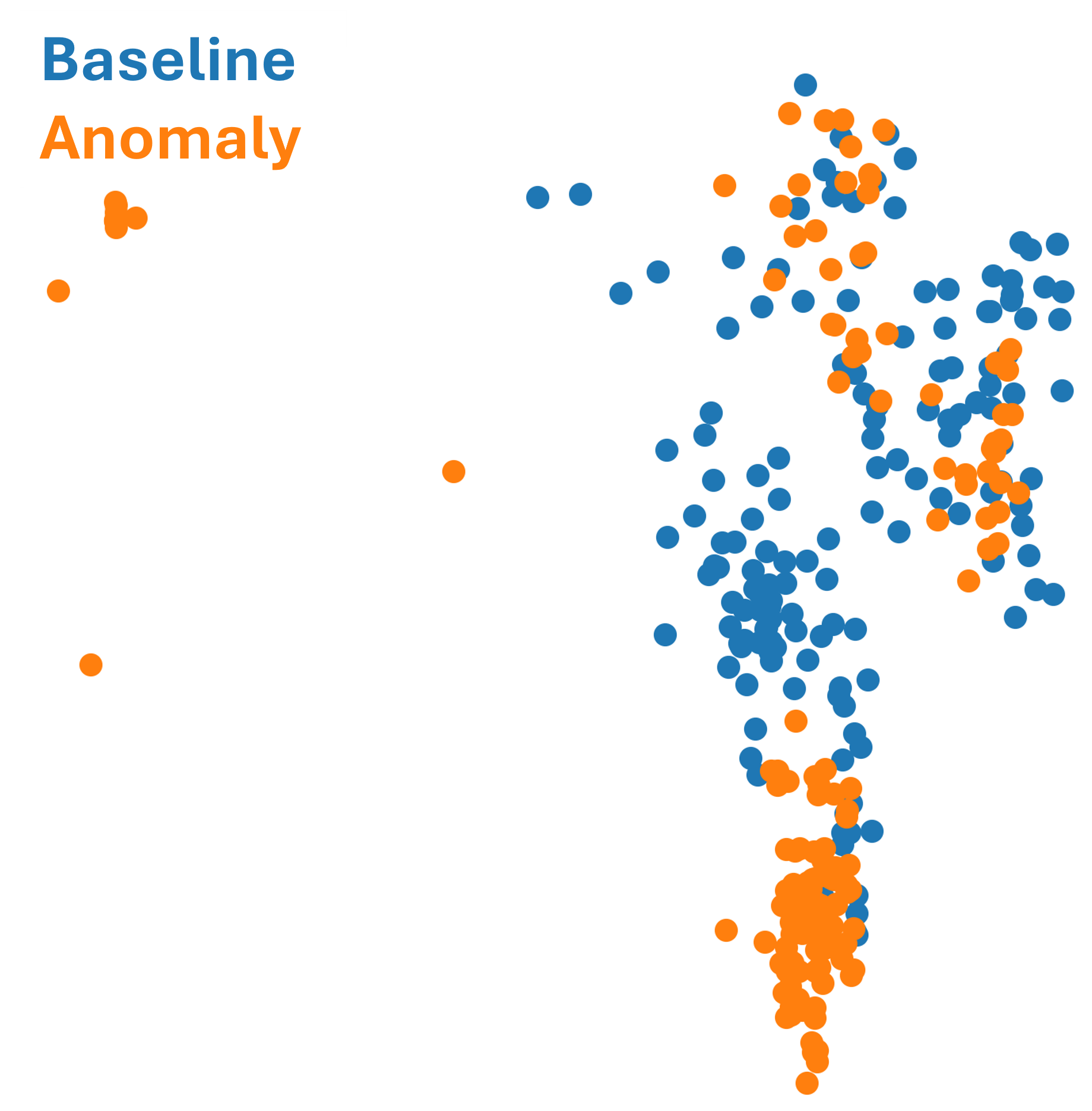}}

    \caption{Quantum Kernel}
    \label{fig:qk}
    \end{subfigure}%
\caption{2D Projection of test kernel matrix with classical and Quantum dot product kernel \textit{blue} represents Baseline or Normal class and \textit{orange} represents Anomaly or Stress class}
\label{fig:2dproj}
\end{figure}

It is seen from Fig.~\ref{fig:2dproj} that, after the kernel computation both classical and quantum kernel grouped the test samples. For classical kernel~\ref{fig:ck} there are a lot of overlaps that make drawing the decision boundary harder for any classifier. On the other hand, with high dimensional correlation the quantum kernel~\ref{fig:qk} has fewer overlaps. The 2D projection thus helped us hypothesize that the quantum SVM can expect better performance.

\section{Result Analysis}
\label{Result}
Our design for the quantum circuit and the simulation was performed using Pennylane~\cite{pennylane} with $qiskit\ aer$ backend~\cite{qiskit} with no noise model. In this section, we describe the experimental result achieved with our stress dataset. We also compare the performance with the equivalent classical model having a similar number of features. To properly identify the benefits of using a fidelity-based quantum kernel, a table of various performance metrics is provided in Table \ref{table:performance}.

\begin{table}[htbp]
\caption{Performance metrics of Quantum SVM model and typical classical SVM based on feature (and qubit) counts. }
\label{table:performance}
\centering
\resizebox{\columnwidth}{!}{
    \setlength{\tabcolsep}{3pt}
    \renewcommand*{\arraystretch}{2}
    \begin{tabular}{cccccc}
    \toprule
     \bf Method & \bf Features & \bf Accuracy & \bf Precision & \bf Recall & \bf F1 \\
        \cmidrule{1-6}
        \multirow{2}{*}{\bf Quantum} & \bf 8 (4-Qubits) & \makecell[c]{Max: 70\%\\Avg: 61\%} & \makecell[c]{Max: 65\%\\Avg: 63\%} & \makecell[c]{Max: 72\%\\Avg: 64\%} & \makecell[c]{Max: 0.75\\Avg: 0.64} \\
        \cmidrule{2-6}
         & \bf 12 (6-Qubits) & \makecell[c]{Max: 79\%\\Avg: 64\%} & \makecell[c]{Max: 80\%\\Avg: 64\%} & \makecell[c]{Max: 79\%\\Avg: 64\%} & \makecell[c]{Max: 0.78\\Avg: 0.63} \\
        \midrule
       % \multirow{4}{*}{\makecell[c]{\bf Quantum \\ \bf (noisy\\ \bf simulation)}} & \bf 8 (4-Qubits) & \makecell[c]{Max: 53\%\\Avg: 47\%} & \makecell[c]{Max: 55\%\\Avg: 49\%} & \makecell[c]{Max: 52\%\\Avg: 46\%} & \makecell[c]{Max: 0.54\\Avg: 0.47} \\
       %  \cmidrule{2-6}
       % & \bf 12 (6-Qubits) & \makecell[c]{Max: 63\%\\Avg: 59\%} & \makecell[c]{Max: 65\%\\Avg: 55\%} & \makecell[c]{Max: 69\%\\Avg: 62\%} & \makecell[c]{Max: 0.67\\Avg: 0.58} \\
       %  \midrule
        \multirow{2}{*}{\bf Classical} & \bf 8 & \makecell[c]{Max: 64\%\\Avg: 57\%} & \makecell[c]{Max: 76\%\\Avg: 59\%} & \makecell[c]{Max: 65\%\\Avg: 59\%} & \makecell[c]{Max: 0.66\\Avg: 0.59} \\
        \cmidrule{2-6}
        & \bf 12 & \makecell[c]{Max: 77\%\\Avg: 62\%} & \makecell[c]{Max: 77\%\\Avg: 64\%} & \makecell[c]{Max: 75\%\\Avg: 65\%} & \makecell[c]{Max: 0.75\\Avg: 0.65} \\
        \bottomrule
    \end{tabular}
    }
\end{table}

The table shows proposed QSVM performed better than classical SVM in all trials. It works best with 12 features and 6 qubits with 79\% accuracy and an F-1 score of 0.78. Again, another observation from the experiment is the low precision and high recall value for quantum SVM. This is particularly, useful for medical diagnosis and anomaly detection where a higher recall value provides an advantage. Thus, the proposed QSVM model proved to be more viable for stress detection. In such a context missing a positive case can be damaging and a higher recall value prevents that. Comparing this to a classical counterpart, the low recall value put the model at a disadvantage.

\section{Conclusion and Future Work}
\label{Discussion}
Our findings demonstrate that by combining the right pre-processing with efficient feature map selection, we can create accurate quantum anomaly detection models for complex data. These findings demonstrate the potential of QML for stress detection in older adults using a wearable watch, signifying its promising application in smart healthcare. However, quantum kernel computation is resource-intensive on existing NISQ machines. Thus, jobs requiring real-time classification, such as stress detection, become significantly more computationally intensive due to the limits of current QPUs. To address this challenge, future research should focus on reducing computational overhead through methods such as circuit parallelization~\cite{Ohkura2022, Domeniconi2001}, hybrid and ensemble methods~\cite{Araya2017}, Quantum Neural Networks~\cite{network}. Additionally, because NISQ computers are prone to noise and error, further study on noise impacts and noise mitigation techniques is essential. 

\section{Acknowledgment}
This research used resources of the Oak Ridge Leadership Computing Facility, which is a DOE Office of Science User Facility supported under Contract DE-AC05-00OR22725.

\bibliographystyle{style}\balance
\bibliography{reference}\balance

\end{document}